# History and presentation of the Cavendish experiment in textbooks in the 20th century

**Jonas R Persson**

Department of Physics, Norwegian University of Science and Technology, NO 7491 Trondheim, Norway

E-mail: jonas.persson@ntnu.no

The experiment performed by Henry Cavendish to measure the density of the earth, is in numerous textbooks described as a measurement of the universal gravitational constant, G, even if we know that this was not true. In this paper, a study on how common this "myth" is based on the checklist developed by Leite on a total of 84 textbooks. The prevalence of the myth in most textbooks throughout the 20th century indicates a focus on the contemporary interests of authors and the physics community in presenting the development of physics. An explanation of the prevalence of the myth and a different approach to presenting the Cavendish experiment in textbooks are suggested.

## 1. Introduction

Throughout the history of physics, numerous groundbreaking experiments have played a pivotal role in shaping the development of fundamental theories. These experiments, often featured in physics textbooks, are not only essential for understanding the underlying principles of physics but also serve as effective didactic tools. They offer insights into new phenomena, the evolution of theories, and the validation of theoretical predictions, all while introducing students to the historical progression of scientific inquiry - an essential aspect of science education [1].

However, presenting these experiments from our modern perspective can sometimes obscure the original motivations and interests of the people performing the experiment, which were shaped by the scientific context and knowledge of their time. By failing to acknowledge these historical perspectives, we risk distorting the true narrative and portraying the scientific development as a linear succession of triumphs. Such a simplified portrayal may inadvertently demotivate students who struggle to grasp the subject, as they might assume that scientific understanding unfolded effortlessly.

It is crucial to recognize that the progress of experiments and theories in physics was influenced not only by scientific interests but also by technological limitations, societal needs, and methodological constraints. For instance, the development of observatory astronomy was closely tied to navigation, while the quest for an accurate chronometer played a pivotal role in solving the Longitude problem [2].

Mathematical and methodological limitations also played a significant role in shaping experiments and measurements. Before the establishment of the current SI system with its traceable base units, scientists used various measurement systems and relied on relative measurements and equations. The transition to dimensionally homogeneous equations, as demonstrated by Joseph Fourier in his work on heat theory [3,4], paved the way for absolute measurements based on standardized units. The pioneering work of Gauss [5] in 1832 enabled the determination of fundamental constants, adding further significance to experiments like the Cavendish experiment.

Considering the above, this paper aims to evaluate how the Cavendish experiment has been presented in physics textbooks over the 20th century. By examining the historical perspectives and comparing them with actual events, I search to understand the disparity between the portrayed history and the true historical narrative. Identifying the sources of incorrect historical descriptions may provide valuable insights into how scientific knowledge is conveyed and evolves in educational settings.

By delving deeper into the historical context of the Cavendish experiment, I hope to uncover a more accurate and nuanced account that showcases both the triumphs and struggles of scientific inquiry. This research aims to enhance science education by presenting a more comprehensive and authentic depiction of the experiments that have shaped the course of physics.

## 2. The Cavendish experiment

The Cavendish experiment holds the distinction of being the first laboratory-scale measurement of the Earth's density. The concept and design of the apparatus were initially conceived by John Michell [6,7], but unfortunately, he was unable to complete the experiment before his passing in 1793.

Michell built a torsion balance, which operated on the same principle as Charles Coulomb's experiments measuring small electric attractions and repulsions. However, it appears that Michell was unaware of Coulomb's work at the time of his death. Michell's primary interest in the experiment was to investigate the validity of Newton's gravitational law on a smaller scale, verifying whether it held not only for celestial bodies but also for objects on Earth.

While the gravitational law had been well-established for astronomical scales, its confirmation on smaller scales remained unexplored. Unfortunately, Michell's apparatus was left unfinished, or he did not achieve the intended accuracy upon his passing. William Wollaston, upon receiving the apparatus, could not carry out the measurements and eventually passed it on to Henry Cavendish.

Cavendish, using Michell's balance as a template, made modifications to improve its accuracy and address issues with design and materials. Cavendish performed a total of 17 measurements using the revised apparatus from August 5th, 1797, to May 30th, 1798. The main objective of the experiment, as explicitly stated in the title of Cavendish's original paper [8], was to determine the density of the Earth.

Detailed descriptions of the experiment can be found in Cavendish's original paper and the articles by Moreno González and Ducheyne [6,7]. The Cavendish experiment, with its innovative approach and groundbreaking results, remains a cornerstone in the history of physics and the determination of the Earth's density.

## 3. Assessment rubric and criteria

The study on the presentation of the Cavendish experiment in textbooks, based on the checklist developed by Leite [9] was conducted in the Physics department, utilizing a comprehensive range of sources, including personal, institutional, and university libraries, as well as digital (scanned) textbooks available from platforms like The Internet Archive [10] and Google Books [11]. A total of 84 textbooks, published between 1911 and 2015, were included in the study. A limited study on the presentation of the Cavendish experiment in textbooks from the first decade of the 21st Century was done by Slisko and Hadzibegovic [12].

The evaluation of the textbooks was based on five criteria related to common textbook presentations, with three coinciding with the criteria in Slisko and Hadzibegovic [12]:

1. Presentation of the experimental setup: The study assessed whether the textbooks presented the experimental setup of the Cavendish experiment through text or figures.
2. Statement about Cavendish measuring/determining Earth's density: The textbooks were checked to see if they explicitly mentioned that Cavendish measured or determined the Earth's density.
3. Statement about Cavendish determining the Newtonian constant of gravitation: The study looked for indications in the textbooks stating that Cavendish determined the Newtonian constant of gravitation, G.
4. Statement about Cavendish determining/measuring the mass of the Earth: The textbooks were examined to determine whether they stated that Cavendish determined or measured the mass of the Earth.
5. Mention of Michell: The study looked for any mention of John Michell, in connection with the Cavendish experiment.

The evaluation applied a conservative judgement in such a way that if the presentation in the textbook could be interpreted as affirmative to the individual questions, it was considered affirmative. This approach was used to avoid potential ambiguity and ensure clarity in the textbooks' presentations, considering that students might make similar interpretations and potentially acquire an incorrect understanding of the material.

By employing these criteria and evaluating the textbooks in this manner, an objective analysis of how the Cavendish experiment and its historical context are portrayed in physics textbooks is hopefully achieved.

## 4. Results of the assessment

The results of the analysis for each criterion are summarised in Table 1 and given in detail in Table 2.

*Table 1 Summary of the evaluation for the different criteria.*

| | Count (Yes) | Count (No) | Count (Inferring Y*) |
|---|---|---|---|
| Criterion 1 | 77 | 7 | |
| Criterion 2 | 2 | 79 | 3 |
| Criterion 3 | 59 | 21 | 4 |
| Criterion 4 | 12 | 60 | 12 |
| Criterion 5 | 13 | 71 | |

### *4.1 CRITERION 1. Description of the experimental setup*

Out of 84 textbooks, seven do not present the experimental setup of the Cavendish experiment. Instead, they describe the law of gravitation from a theoretical perspective and mention the Cavendish experiment and Cavendish himself. For example:

"*The constant G that appears in Equation 4.3 is called the universal gravitational constant, because it has the same value for all pairs of particles anywhere in the universe, no matter what their separation. The value of G was first measured in an experiment by the English scientist Henry Cavendish (1731-1810), more than a century after Newton proposed his law of universal gravitation."* [13,p.96]

This and other omissions might be due to streamlining the presentation or considering the Cavendish experiment as old-fashioned or outdated, as these textbooks were published in the last 35 years (1991-2010).

### *4.2 CRITERION 2. Cavendish measured the earth's density.*

Only five out of 84 textbooks state in a clear and definite way that Cavendish's primary goal was to measure the density of the Earth. Some textbooks mention the density as a secondary result derived from experimental values of the mass of the Earth [14] or from determining G [15-17].

One also find ambiguous presentations, as shown in the quote from [18]:

"*With these experiments Cavendish determined G. He expressed his answer in terms of the mean density of the Earth, which he found to be close to five and a half times that of water, very close to Newton's estimate"* [18, p. 360]

Even if the presentation seems to point to a determination of the earth's density, the statement that Cavendish determined G could be considered as not fulfilling the criterion as the focus is on G and not the density of the earth. It might be a misunderstanding or a mix of different concepts that were not resolved in the preparation of the manuscript. From a student's view, the statement will be confusing.

The best description of the actual events is given by Serway, Jewett et al. [19]:

"*The universal gravitational constant G was first evaluated in the late 19th century, based on results of an important experiment by Sir Henry Cavendish in 1798.... Cavendish's goal was to measure the density of the Earth. His results were then used by other scientists 100 years later to generate a value for G.*" [19, p. 389]

It should be noted that some textbooks are ambiguous when it comes to the description of the results. For example, Resnick, Halliday et al. [20,21] first state:

"*The first laboratory determination of G from the force between spherical masses at a close distance was done by Henry Cavendish in 1798.*" [20, p346, 21, p303]

Later in the presentation, the authors seem to back down from this claim with a sentence in parenthesis:

"*In fact, the title of the paper written by Cavendish describing his experiments referred not to measuring G but instead to determining the density of the earth from its weight and volume*". [20, p347, 21, p303]

Even if they provide the reader with an erroneous correction, the textbooks do not fulfil the criterion, as they talk about the determination through the weight and radius of the earth. This indicates that the authors may very well have been aware of the history but chose to follow the presentation in other textbooks.

### *4.3 CRITERION 3. Cavendish determined G.*

Due to the ambiguity in the presentations, it is sometimes difficult to know if the textbooks discuss if the determination of G was done by Cavendish or if his results were used by others. The interpretation of whether this is inferred or just due to bad formulation is open for discussion and is out of the scope of this paper. I have chosen to interpret a discussion on Cavendish and G in such a way that if the discussion can be misunderstood to be affirmative to Cavendish determining G, then this criterion is fulfilled. In this way, I place myself in the position of students who might not read the text in detail or reflect on it. Still, 63 of the textbooks give direct or indirect the impression that Cavendish determined G experimentally.

The possible ambiguity of the presentations is exemplified in the following quotes:

"*In order to make quantitative predictions and analysis of physical phenomena involving gravitational interactions, it is necessary to know the universal gravitational constant G. In 1797-1798 Henry Cavendish performed the first experiment to determine a precise value for G.*" [22, p 122]

The authors seem to indicate that Cavendish did the experiment aiming at determining G.

"*In 1798, over 100 years later, Cavendish measured the gravitational interaction for laboratory-sized objects and*

*calculated from his observations the value of the constant G.*" [23, p61]

The author correctly describes the experiment as being on the gravitational interaction but claims that Cavendish calculated a value of G.

"*To find G it is necessary to measure the force with which two bodies of known masses attract other when they are at a known distance apart. This was first done by Cavendish about the year 1797.*" [24, p 100]

This quote can be interpreted as referring to the Cavendish experiment on the gravitational interaction and not to the determination of G. Still, a superficial reader will probably interpret this as a determination of G rather than a measurement of the gravitational interaction.

In most cases, one finds clear statements that Cavendish determined G.

"*The value of G was first determined by Henry Cavendish in 1798....*" [25, p 323]

"*The universal gravitational constant G was first measured in 1798 by Henry Cavendish, who used the apparatus shown in figure....*" [26, p.345]

The presentations do not show any major difference between different decades, indicating that this belief originates early in the history of describing the determination of G.

In addition, one find some textbooks that do not give the impression that Cavendish measured G or the density of the Earth. The focus is more on the method or apparatus used, something that can be called an engineering or technical approach. An example of this is given in Sears and Zemansky's University Physics [27]:

"*The numerical value of the constant G depends on the units in which force, mass and distance are expressed. Its magnitude can be found experimentally by measuring the force of gravitational attraction between two bodies of known masses m and m', at a known separation. For bodies of moderate size the force is extremely small, but it can be measured with an instrument that which was invented by the Rev. John Michell, although it was first used for this purpose by Sir Henry Cavendish in 1798. The same type of instrument was also used by Coulomb for studying forces of electrical and magnetic attraction and repulsion.*" [27, p. 79]

A formulation that stays the same until the 7th edition. This is also in line with the aim formulated by the authors in the first editions with a focus on physics and methods. It is also interesting to note the drastic change with the 8th edition [28] in 1992:

"*To determine the value of the gravitational constant G we have to measure the gravitational force between two bodies of known masses* $m_1$ *and* $m_2$ *at a known distance r. For bodies of reasonable size the force is extremely small, but it can be measured with an instrument called torsion balance, used by Sir Henry Cavendish in 1798 to determine G.*" [28, p. 317]

The eighth edition [28] also introduced a change in the philosophy of introductory physics courses as stated in the preface. The aim was to make physics more human with in some cases more historical background. However, as shown by Persson [29] in the case of black-body radiation, this included an incorrect history, something that is seen here as well.

### *4.4 CRITERION 4. Cavendish determined the mass of the earth.*

The Cavendish experiment has also been described as the experiment that weighed the earth, so it is also interesting to see how common this belief is. I found this in 24 out of 84 textbooks, some with the claim that Cavendish "weighed the earth". As this is a description of the experiment in popular presentations, this idea also seems to have permeated into textbooks. In some respect this is a better description of the experiment than the determination of G, but that was not the aim of the experiment as with criterion 3. This claim does not seem to be dependent on when the textbook was published, an indication of an early introduction of this belief.

### *4.5 CRITERION 5. Mention of Michell.*

Only 13 textbooks acknowledge John Michell, the original designer of the apparatus used in the Cavendish experiment. His contribution to the experiment and its background are not widely mentioned, and he seems to have been forgotten in most cases.

Sears and Zemansky mention Michell in six editions of University Physics [27]. One also find that Michell is mentioned in 4 books from the beginning of the 20th century, but otherwise, he seems to have been forgotten.

The study reveals inconsistencies and ambiguities in the presentation of the Cavendish experiment and its historical context in physics textbooks in the same manner as was done by Slisko and Hadzibegovic [12] for a limited number of textbooks. Some textbooks accurately describe Cavendish's primary goal as measuring the density of the Earth, while others focus more on his determination of G or the mass of the Earth. Additionally, John Michell's contribution is often overlooked despite being the original designer of the apparatus used in the experiment.

## 5. Origins of Historical Inaccuracy in the Determination of the Gravitational Constant by Cavendish

Examination of textbooks reveals an early origin of the misrepresentation surrounding Cavendish's purported determination of the gravitational constant (G). Even in the earliest textbook, an erroneous attribution to Cavendish is evident, wherein authors mistakenly credit him with establishing the value of G. Remarkably, the 1911 edition of the Encyclopaedia Britannica [30, Vol V, p581] offers a precise description of the experiment, duly acknowledging John Michell's contributions and the experiment's primary objective.

The Encyclopaedia Britannica of 1911 articulates, "Cavendish's last great achievement was his famous series of experiments to determine the density of the earth. The apparatus he employed was devised by the Rev. John Michell..." [30, Vol V, p581] This accurate representation persists in contemporary iterations of the Encyclopaedia Britannica and Wikipedia.

It is important to underscore that the gravitational constant did not command scientific attention until the latter part of the 19th century. The first derivation of the gravitational constant I found was in a publication by Cornu and Baille [31] in 1873, with recognition of its importance by Everett in 1875 [32] and Maxwell in "A Treatise on Electricity and Magnetism" [33, p4] as a way to define an astronomical mass unit.

The fallacious notion that Cavendish determined G appears to have its origin during the last decades of the 19th century, coinciding with a scientific paradigm shift from an emphasis on Earth's density to an inquiry into the gravitational constant. The earliest explicit attribution of a determination of G to Cavendish is discerned in C.V. Boys' paper, "On the Cavendish experiment" [34], presented at the Royal Society in June 1889. Boys, doing experiments directed at the gravitational constant, inadvertently presented a narrative more based on his research interest than on history. In it, he claims "*the Cavendish experiment for determining the constant gravitation from which the density of the earth may be calculated is so well known that there is no occasion to describe it.* " [34]

Noteworthy from the same period is the writings of J.H. Poynting where the density of the earth is more important. In his essay "The Mean Density of the Earth" [35], there is an absence of any indication that Cavendish's pursuits extended beyond a determination of the Earth's density. One must also be aware that Poynting published a paper in 1891 [36] on his experiment to determine the density of the Earth and the gravitational constant, so he was aware of the history of the Cavendish experiment. Poynting's experiment was first set up in the Cavendish laboratory through the kindness of Prof. Maxwell, which might indicate an interest from Maxwell's side in a better determination of the gravitational constant at the time of the first attempts by Poynting.

The provenance of the conviction that Cavendish determined G can be traced back to Boys' 1889 Royal Society presentation, thereby shaping subsequent narratives characterizing the Cavendish experiment as principally concerned with establishing the gravitational constant rather than elucidating the density of the Earth. It merits attention that Boys' presentation found dissemination in the popular press, akin to the historical narrative analyses conducted by Passon [37] elucidating the origins of Kelvin's clouds and the UV catastrophe.

Based on this it's quite likely that the origin of belief that Cavendish determined G, originates from C.V. Boys' presentation in 1889 at the Royal Society and that this presentation became the starting point in describing the Cavendish experiment as an experiment of determining the gravitational constant and not the density of the Earth. One should also note that Boys' presentation was reported in the popular press, something that might have further contributed to the spread of this narrative.

## 6. Discussion

Textbooks are expected to provide accurate descriptions of scientific concepts, but when it comes to the history of science, this is not always the case. Additionally, textbooks serve as a means to present current scientific practices and reflect the ideology of science. It is crucial to recognize that scientific history presented in textbooks often highlights the successful outcomes and scientific methods of "victors," while overlooking mistakes and dead ends.

The historical narrative surrounding the Cavendish experiment is a prime example of how textbooks focus on themes considered important in contemporary scientific contexts. The fundamental gravitational constant G takes precedence over the experiment's more practical objective of determining the Earth's density, which becomes less interesting in today's perspective.

The contextual nature of studying textbook presentations is essential to understand the choices made by authors. Textbook content reflects the prevailing image and current interests of science. This realization leads to the inference that the emphasis on fundamental constants like G became dominant at the beginning of the 20th century, evident from the treatment of the Cavendish experiment in textbooks from that period.

Encyclopaedias, not typically authored by physicists, are less influenced by contemporary scientific contexts. As a result, they often provide more accurate and diverse historical representations.

Efforts in the 1980s to humanize physics and present it in a more relatable context demonstrate attempts to introduce changes in textbook narratives. However, even in these cases, the historical information may still be based on previous physics textbooks rather than original sources or works by historians of science, leading to a perpetuation of certain ideas.

In conclusion, recognizing the contextual nature of textbook presentations is crucial for a nuanced understanding of the history of science and how it is presented. Critically evaluating textbook content and seeking diverse sources of historical information allow educators and students to develop a comprehensive view of the evolution of scientific concepts and experiments throughout history.

## 7. Conclusion and implications

The study has revealed that the false history of the Cavendish experiment has prevailed in physics textbooks since the early 20th century. The likely origin of this false narrative can be traced back to C.V. Boy's presentation of his measurements of the gravitational constant in 1889 [34] and subsequent scientific and popular reports. This highlights the importance of critically examining historical sources and their impact on shaping the narrative presented in textbooks.

To present the proper history of the Cavendish experiment, it is not only essential for it to align with the scientific context but also to serve the present ideology of science. In other words, the historical account must be relevant to how science is practised today. Simply highlighting the experiment's aim to determine the density of the Earth or its role in confirming gravitational interaction at small distances may not be sufficient to correct the false history in textbooks.

However, by exploring why Cavendish could not determine G accurately, an opportunity arises to provide a more nuanced and comprehensive historical account. It becomes relevant to discuss the development of unit systems and the importance of using homogeneous dimensional equations, which played a significant role in shifting the focus toward fundamental constants during the 19th century. By connecting the Cavendish experiment with the evolution of unit systems and absolute measurements, the proper history can be reinstated in physics textbooks.

This approach serves a dual purpose: it rectifies the historical inaccuracy while conveying the importance of standardised unit systems and dimensional homogeneous equations in the context of scientific progress. By contextualizing history in this manner, textbooks can present a more accurate and insightful narrative, providing students with a deeper understanding of the scientific principles and the historical context in which they were developed.

Table 2. Evaluation of Textbooks based on the presentation of the Cavendish experiment. Y criterion affirmative, N criterion negative, * inferring.

| Textbook | Criteria1 Experimental set-up | Criteria2 Cavendish determined Earth's density | Criteria 3 Cavendish determined G | Criteria 4 Cavendish determined Earth's mass | Criteria 5 Mentions Michell | Notes |
|---|---|---|---|---|---|---|
| Edser 1911 | **Y** | **Y*** | **Y** | **N** | **Y** | |
| Watson 1911 | **Y** | **N** | **Y*** | **N** | **Y** | |
| Anderson 1914 | **Y** | **Y*** | **N** | **Y** | **N** | |
| Wilson 1915 | **Y** | **N** | **Y** | **N** | **N** | Quote referred to in the text. |
| Duff, Lewis et al. 1916 | **Y** | **N** | **Y** | **Y*** | **N** | |
| Duff, Carman et al. 1921 | **Y** | **N** | **Y** | **Y*** | **N** | |
| Ferry 1921 | **Y** | **N** | **N** | **N** | **Y** | |
| Hering 1921 | **Y** | **N** | **Y*** | **Y*** | **Y** | |
| Duff, Lewis et al. 1937 | **Y** | **N** | **Y** | **Y*** | **N** | |
| Smith 1938 | **Y** | **N** | **Y** | **N** | **N** | |
| Eldridge 1940 | **Y** | **N** | **N** | **N** | **N** | |
| Robeson 1942 | **Y** | **N** | **Y** | **N** | **N** | |
| Black 1948 | **Y** | **N** | **N** | **N** | **N** | |
| Sears and Zemansky 1949 | **Y** | **N** | **N** | **N** | **N** | 1st ed. |
| Sears 1952 | **Y** | **N** | **N** | **N** | **N** | |
| Margenau 1953 | **Y** | **N** | **Y** | **Y** | **N** | |
| Rusk 1954 | **Y** | **N** | **Y** | **N** | **Y** | |
| Sears and Zemansky 1955 | **Y** | **N** | **N** | **N** | **Y** | 2nd ed. Quote referred to in the text. |
| Shortley and Williams 1955 | **Y** | **N** | **Y** | **N** | **N** | |
| Semat and Katz 1958 | **Y** | **N** | **Y** | **N** | **N** | |
| Physical Science Study Committee. 1960 | **Y** | **Y*** | **Y** | **N** | **N** | Quote referred to in the text. |
| Resnick and Halliday 1960 | **Y** | **N** | **Y** | **Y*** | **N** | |
| Beiser 1962 | **Y** | **N** | **N** | **N** | **N** | |
| Sears and Zemansky 1963 | **Y** | **N** | **N** | **N** | **Y** | 3rd ed. |
| Alonso and Finn 1967 | **Y** | **N** | **N** | **N** | **N** | |
| Cooper 1968 | **Y** | **N** | **Y** | **N** | **N** | Quote referred to in the text. |
| Bueche 1969 | **Y** | **N** | **Y** | **N** | **N** | |
| McCormick 1969 | **Y** | **N** | **Y** | **N** | **N** | |
| Sears and Zemansky 1970 | **Y** | **N** | **N** | **N** | **Y** | 4th ed. |
| Borowitz and Beiser 1971 | **Y** | **N** | **Y** | **N** | **N** | |
| Harvard Project Physics. 1971 | **Y** | **N** | **Y*** | **N** | **N** | |
| Marion 1975 | **Y** | **N** | **Y** | **N** | **N** | |
| Sears, Zemansky et al. 1976 | **Y** | **N** | **N** | **N** | **Y** | 5th ed. |
| Resnick and Halliday 1977 | **Y** | **N** | **Y** | **Y*** | **N** | |
| Orear 1979 | **Y** | **N** | **Y** | **Y** | **N** | |
| Physical Science Study Committee., Haber-Schaim et al. 1981 | **Y** | **N** | **Y** | **N** | **N** | |
| Sears, Zemansky et al. 1982 | **Y** | **N** | **N** | **N** | **Y** | 6th ed. |
| Ohanian 1985 | **Y** | **N** | **N** | **N** | **N** | |
| Blatt 1986 | **Y** | **N** | **Y** | **Y*** | **N** | |
| Sears, Zemansky et al. 1987 | **Y** | **N** | **N** | **N** | **Y** | 7th ed. |
| Halliday and Resnick 1988 | **Y** | **N** | **Y** | **N** | **N** | |
| Blatt 1989 | **Y** | **N** | **Y** | **Y*** | **N** | |
| Cutnell and Johnson 1989 | **Y** | **N** | **Y** | **N** | **N** | |
| Gettys, Keller et al. 1989 | **Y** | **N** | **Y** | **Y*** | **N** | |
| Giancoli and Giancoli 1989 | **Y** | **N** | **Y** | **Y** | **N** | |
| Ohanian 1989 | **Y** | **N** | **N** | **N** | **N** | |
| Serway 1990 | **Y** | **N** | **Y** | **N** | **N** | |
| Sternheim and Kane 1991 | **N** | **N** | **Y** | **Y** | **N** | |
| Tipler and Tipler 1991 | **Y** | **N** | **Y** | **N** | **Y** | |
| Resnick, Halliday et al. 1992 | **Y** | **N** | **Y** | **Y*** | **N** | Quote referred to in the text. |
| Young and Sears 1992 | **Y** | **N** | **Y** | **Y** | **N** | 8th ed. Change of narrative and Michell omitted. Quote referred to in the text. |
| Keller, Gettys et al. 1993 | **Y** | **N** | **Y** | **Y*** | **N** | |
| Bueche and Jerde 1995 | **Y** | **N** | **Y** | **N** | **N** | |
| Coletta 1995 | **Y** | **N** | **Y** | **N** | **N** | |
| Fishbane, Gasiorowicz et al. 1996 | **Y** | **N** | **Y** | **N** | **N** | Quote referred to in the text. |
| Lerner 1996 | **Y** | **Y** | **Y*** | **N** | **Y** | |
| Serway 1996 | **Y** | **N** | **Y** | **N** | **N** | |
| Mansfield and | **N** | **N** | **Y** | **N** | **N** | |
| O'Sullivan 1998 | | | | | | |
| Tipler 1999 | **Y** | **N** | **Y** | **N** | **N** | |
| Reese 2000 | **Y** | **N** | **Y** | **N** | **N** | |
| Serway, Beichner et al. 2000 | **Y** | **N** | **Y** | **N** | **N** | |
| Young, Freedman et al. 2000 | **Y** | **N** | **Y** | **Y** | **N** | $10^{th}$ ed. |
| Grant and Phillips 2001 | **Y** | **N** | **N** | **N** | **N** | |
| Cassidy, Holton et al. 2002 | **Y** | **N** | **Y** | **N** | **N** | |
| Resnick, Halliday et al. 2002 | **Y** | **N** | **Y** | **Y*** | **N** | Quote referred to in the text. |
| Walker 2002 | **Y** | **N** | **Y** | **Y** | **N** | |
| Cummings, Halliday et al. 2004 | **N** | **N** | **N** | **N** | **N** | |
| Tipler and Mosca 2004 | **Y** | **N** | **Y** | **N** | **N** | Quote referred to in the text. |
| Knight 2004 | **Y** | **N** | **Y** | **N** | **N** | |
| Fishbane, Gasiorowicz et al. 2005 | **Y** | **N** | **Y** | **N** | **N** | |
| Giancoli 2005 | **Y** | **N** | **Y** | **Y** | **N** | |
| Serway and Jewett 2006 | **Y** | **N** | **Y** | **N** | **N** | |
| Touger 2006 | **Y** | **N** | **Y** | **N** | **N** | |
| Cutnell and Johnson 2007 | **N** | **N** | **Y** | **N** | **N** | |
| Ohanian and Markert 2007 | **Y** | **N** | **N** | **N** | **N** | |
| Walker 2007 | **Y** | **N** | **Y** | **Y** | **N** | |
| Wilson, Buffa et al. 2007 | **N** | **N** | **Y** | **N** | **N** | |
| Giancoli 2008 | **Y** | **N** | **Y** | **Y** | **N** | |
| Young, Freedman et al. 2008 | **Y** | **N** | **Y** | **Y** | **N** | 12th ed. |
| Cutnell, Johnson et al. 2010 | **N** | **N** | **Y** | **N** | **N** | Quote referred to in the text. |
| Mansfield and O'Sullivan 2010 | **N** | **N** | **N** | **N** | **N** | |
| Chabay and Sherwood 2011 | **Y** | **N** | **Y** | **N** | **N** | |
| Serway, Jewett et al. 2014 | **Y** | **Y** | **N** | **N** | **N** | Quote referred to in the text. |
| Chabay and Sherwood 2015 | **Y** | **N** | **Y** | **N** | **N** | Quote referred to in the text. |

Table 3 List of books analysed.

| |
|---|
| Alonso, M., and E. J. Finn (1967). Fundamental university physics. Reading, Mass., Addison-Wesley Pub. Co. |
| Anderson, W. B. (1914). Physics for technical students. New York, McGraw-Hill book company, inc.; etc. |
| Beiser, A. (1962). The mainstream of physics. Reading, Mass., Addison-Wesley Pub. Co. |
| Black, N. H. (1948). An introductory course in college physics. New York, Macmillan Co. |
| Blatt, F. J. (1986). Principles of physics. Boston, Allyn, and Bacon. ISBN 0-205-08555-5 |
| Blatt, F. J. (1989). Principles of physics. Boston, Allyn, and Bacon. ISBN 0-205-11784-8 |
| Borowitz, S. and A. Beiser (1971). Essentials of Physics; a text for students of science and engineering. Reading, Mass., Addison-Wesley Pub. Co. |
| Bueche, F. (1969). Introduction to physics for scientists and engineers. New York, McGraw-Hill. |
| Bueche, F. J. and D. A. Jerde (1995). Principles of physics, McGraw-Hill. ISBN 0-07-008817-9 |
| Cassidy, D. C., et al. (2002). Understanding physics. New York, Springer. ISBN 0-387-98756-8 |
| Chabay, R. W. and B. A. Sherwood (2011). Matter & interactions. Hoboken, NJ, Wiley. ISBN 978-0-470-50347-8 |
| Chabay, R. W. and B. A. Sherwood (2015). Matter & interactions. Hoboken, NJ, John Wiley & Sons. ISBN 978-1-118-87586-5 |
| Coletta, V. P. (1995). College physics. St. Louis, Mosby. ISBN 0-8016-7722-X |
| Cooper, L. N. (1968). An introduction to the meaning and structure of physics. New York, Harper & Row. |
| Cummings, K., et al. (2004). Understanding physics. Hoboken, NJ, Wiley. |
| Cutnell, J. D. and K. W. Johnson (1989). Physics. New York, Wiley. ISBN 978-0-471-61719-8 |
| Cutnell, J. D. and K. W. Johnson (2007). Physics. Hoboken, NJ, Wiley. ISBN 978-0-471-66315-7 |
| Cutnell, J. D., et al. (2010). Introduction to physics. Hoboken, NJ, Wiley. ISBN 978-0-470-40942-8 |
| Duff, A. W., et al. (1921). A text-book of physics. Philadelphia, P. Blakiston's sons & co. |
| Duff, A. W., et al. (1937). Physics for students of science & engineering: Mechanics and sound. Philadelphia, Blakiston. |
| Duff, A. W., et al. (1916). A text-book of physics. Philadelphia, P. Blakiston's son & co. |
| Edser, E. (1911). General physics for students: A text-book on the fundamental properties of matter. London, Macmillan, and Co., limited. |
| Eldridge, J. A. (1940). College physics. New York, London, J. Wiley & sons Chapman & Hall, limited. |
| Ferry, E. S. (1921). General physics and its application to industry and everyday life. New York, John Wiley & sons, inc.; etc. |